\documentclass[aps,prl,twocolumn,superscriptaddress,showpacs]{revtex4}
\usepackage{hyperref}

\usepackage{graphicx}

\usepackage{amsmath}
\usepackage{epsfig}

\begin{document}
\title{Experimental distillation of squeezing from non-Gaussian quantum states}

\author{J. Heersink}
\affiliation{Institut f\"{u}r Optik, Information und Photonik, Max-Planck Forschungsgruppe, Universit\"{a}t Erlangen-N\"{u}rnberg, G\"{u}nther-Scharowsky-Str. 1, 91058, Erlangen, Germany}
\email{joel.heersink@optik.uni-erlangen.de}
\author{Ch. Marquardt}
\affiliation{Institut f\"{u}r Optik, Information und Photonik, Max-Planck Forschungsgruppe, Universit\"{a}t Erlangen-N\"{u}rnberg, G\"{u}nther-Scharowsky-Str. 1, 91058, Erlangen, Germany}
\author{R. Dong}
\affiliation{Institut f\"{u}r Optik, Information und Photonik, Max-Planck Forschungsgruppe, Universit\"{a}t Erlangen-N\"{u}rnberg, G\"{u}nther-Scharowsky-Str. 1, 91058, Erlangen, Germany}
\author{R. Filip}
\affiliation{Institut f\"{u}r Optik, Information und Photonik, Max-Planck Forschungsgruppe, Universit\"{a}t Erlangen-N\"{u}rnberg, G\"{u}nther-Scharowsky-Str. 1, 91058, Erlangen, Germany}
\affiliation{Department of Optics, Palack\'y University, 17. listopadu 50, 77200 Olomouc, Czech Republic}
\author{S. Lorenz}
\affiliation{Institut f\"{u}r Optik, Information und Photonik, Max-Planck Forschungsgruppe, Universit\"{a}t Erlangen-N\"{u}rnberg, G\"{u}nther-Scharowsky-Str. 1, 91058, Erlangen, Germany}
\author{G. Leuchs}
\affiliation{Institut f\"{u}r Optik, Information und Photonik, Max-Planck Forschungsgruppe, Universit\"{a}t Erlangen-N\"{u}rnberg, G\"{u}nther-Scharowsky-Str. 1, 91058, Erlangen, Germany}
\author{U.L. Andersen}
\affiliation{Institut f\"{u}r Optik, Information und Photonik, Max-Planck Forschungsgruppe, Universit\"{a}t Erlangen-N\"{u}rnberg, G\"{u}nther-Scharowsky-Str. 1, 91058, Erlangen, Germany}

\date{\today}

\begin{abstract}
We show theoretically and experimentally that single copy distillation of squeezing from continuous variable non-Gaussian states is possible using linear optics and conditional homodyne detection. A specific non-Gaussian noise source, corresponding to a random linear displacement, is investigated. Conditioning the signal on a tap measurement, we observe probabilistic recovery of squeezing.
\end{abstract}

\pacs{03.67.-a, 42.50.Dv}

\maketitle

Non-classical states such as continuous variable (CV) entangled and squeezed states serve as enabling resources for many CV quantum information protocols~\cite{braunstein05.rmp} as well as for highly sensitive measurements beyond the shot noise limit~\cite{giovannetti04.sci}. The efficiency of these applications relies crucially on the state's nonclassicality (i.e. the degree of single- or two-mode squeezing). Therefore, uncontrolled and unavoidable interaction of the system with the environment and the resultant loss of squeezing in generation or transmission should be combated. This can be done by using a distillation protocol which probabilistically selects out squeezed states from a mixture, hereby increasing the output state's squeezing.

Various protocols exploiting non-Gaussian operations to probabilistically distill two-mode squeezed {\it Gaussian states} have been proposed \cite{duan00.prl}. These protocols are, however, experimentally challenging. In addition it has been proven that the distillation of two-mode squeezed Gaussian states by means of more feasible local Gaussian operations is impossible~\cite{giedke02.pra}. Similarly, following a set of simple arguments, we conjecture that the single copy distillation of single-mode Gaussian squeezed states is impossible using only linear optics and homodyne detection.

There has however been no work devoted to the distillation of CV Gaussian states corrupted by {\it non-Gaussian noise}. This occurs naturally in channels with fluctuating properties, i.e. gain or phase, examples of which are the fading channel~\cite{haas96.i3esac} or channels producing mixture noise~\cite{middleton77.i3etec}. Recently, non-Gaussian telegraph noise has been discussed in qubit systems~\cite{gutmann05.pra}. Therefore, extending the work on Gaussian noise, we pose the question: Is it possible to distill single-mode Gaussian squeezed states with superimposed {\it non-Gaussian noise} using linear optics and homodyne detectors? We answer this question in the affirmative and provide an experimental demonstration.

The set of non-Gaussian noise sources is large. Thus we restrict our attention to a specific case: a squeezed vacuum state perturbed by phase kicks or jitter, attributable to either imperfect generation or transmission through a noisy channel. Assuming these perturbations cause a linear phase space displacement, a convex mixture of two Gaussian squeezed states is created
\begin{equation}\label{W}
W(x,p)=(1-\gamma) W_0(x,p)+\gamma W_1(x,p),
\end{equation}
where $\gamma$ is the probability of displacement, and the individual constituents of the mixture ($i=0,1$) are described by the Wigner functions
\begin{equation}\label{sq}
W_i(x,p)= \frac{ \exp \left(-\frac{(x-\bar{x}_i)^2}{2 \Delta^2 X_{sq}} -\frac{(p-\bar{p}_i)^2}{2 \Delta^2 P_{sq}}\right) }  {2\pi\sqrt{\Delta^2 X_{sq} \Delta^2 P_{sq}}} .\nonumber
\end{equation}
Here $x$ and $p$ are the amplitude and phase quadratures. $\Delta^2 X_{sq}$ and $\Delta^2 P_{sq}$ are the corresponding variances of the input state. $\bar{x}_0,\bar{p}=0$ and $\bar{x}_1,\bar{p}_1$ are the mean values of the initial and displaced squeezed states respectively. We assume the two individual Gaussians to be equally squeezed in $x$: $\Delta^2 X_{sq}<1$, where $\Delta^2 X_{sq} \Delta^2 P_{sq} \geq 1$. The first two moments of the amplitude quadrature of $W(x,p)$ are $\langle x \rangle=\gamma \bar{x}_1$ and $\langle x^2 \rangle= \Delta^2 X_{sq}+\gamma \bar{x}^2_1$, and thus the variance of the amplitude quadrature of the corrupted state of Eq.~\ref{W} is $\Delta^2 X = \Delta^2 X_{sq}+\gamma(1-\gamma)\bar{x}_{1}^{2}$. The second term here originates from the noise and degrades the squeezing. The aim is to recover the squeezing by distilling the initial squeezed state from this mixture. 

\begin{figure}
\includegraphics[scale=0.49]{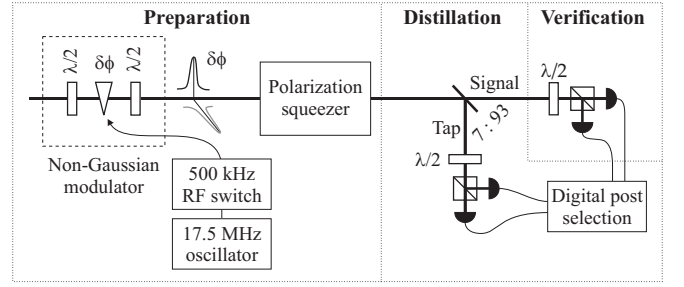}
\caption{Schematic of the experimental setup for the generation, distillation and verificaiton of non-Gaussian squeezed states.}
\label{setup}
\end{figure}

A schematic of the distillation protocol is shown in Fig.~\ref{setup}. A polarization squeezed state, mathematically equivalent to a squeezed vacuum state~\cite{josse03.prl,heersink05.ol}, is modulated to generate a noisy non-Gaussian state. This is incident upon a beam splitter with reflection $R$ and transmission $T$, producing correlated output states. Using a Stokes, or equivalently homodyne, detector a given quadrature of the tap beam is measured. Conditioned on this measurement, the signal is selected only if the outcome lies above a given threshold value, as in Ref.~\cite{laurat03.prl}. Due to the correlations between the signal and tap beams, the scheme accomplishes a probabilistic distillation of the noisy input state. A similar strategy was proposed to purify decohered Schr\"odinger cat states~\cite{suzuki05.xxx}.

We now present a theoretical description of the distillation. The tap beam splitter transforms the quadratures of the Wigner function to, for the transmitted signal beam, $x_s=\sqrt{T}x+\sqrt{R}x_v$ and $p_s=\sqrt{T}p+\sqrt{R}p_v$ where $x_v$ and $p_v$ are uncorrelated vacuum contributions. We write the signal Wigner function after i) detection of the tapped signal and ii) post selection of the signal as
\begin{eqnarray}
W(x_s,p_s)&=&\frac{1}{\Pi}\left[(1-\gamma) G_0(x_s,p_s)W_0(x_s,p_s)+\right. \nonumber \\
& &\left. \gamma G_1(x_s,p_s)W_1(x_s,p_s)\right],
\end{eqnarray}
where $\Pi$ is the success probability and $G_i(x_s,p_s)$ is a filter function which incorporates the effect of the tap measurement and post selection. It thus depends on the measured quadrature and the threshold value. Since the goal of the distillation is to recover the initial squeezing, we consider only the marginal quadrature distribution associated with the squeezed quadrature, $x$. Measuring the phase quadrature in the tapped signal $p_t$~\cite{note}, the resulting probability distribution of the squeezed quadrature in the signal reads
\begin{equation}\label{P}
P(x_s)=\frac{1}{\Pi}\left\{(1-\gamma) g_0P_0(x_s)+\gamma g_1P_1(x_s)\right\},
\end{equation}
where $\Pi=(1-\gamma)g_0+\gamma g_1$ and the individual marginals $P_0(x_s)$ and $P_1(x_s)$ are Gaussian functions with variance $\Delta^2 X_s =T \Delta^2 X_{sq} + R \Delta^2 X_v$ and centered at $\bar{x}_0=0$ and $\sqrt{T}\bar{x}_1$, respectively. The filter function in Eq.~\ref{P} is
\begin{equation}
g_i=\frac{1}{2}\mbox{Erfc}\left[\frac{p_{th}-\bar{p}_i\sqrt{R}}{\sqrt{2 \Delta^2 P_t}} \right].\nonumber
\end{equation}  
Here $p_{th}$ is the post selection threshold and $\Delta^2 P_t=R \Delta^2 P_{sq} + T \Delta^2 P_v$. After distillation the first two moments of the signal are $\langle x_s\rangle=(\sqrt{T}\bar{x}_{1})/(1+r)$ and $\langle
x_s^2\rangle=\Delta^2 X_s+(T\bar{x}_{1}^{2})/(1+r)$ where $r=(1-\gamma)g_0/\gamma g_1$. Thus the distilled squeezing is given by
\begin{equation}
\Delta^2 X_s^{\mathit{distill}}=\Delta^2 X_s+T\bar{x}_1^2\frac{r}{(1+r)^2}.
\label{distill} \end{equation}
The signal variance can be decreased or even the squeezing recovered by minimizing the second term. The probability $\gamma$ and the displacement $\bar{x}_1$ are parameters of the noisy process and thus can not be altered in the distillation optimization. However through the choice of the threshold value $p_{th}$, the ratio between the filter functions $r$ can be controlled to yield efficient distillation, corresponding to $r\rightarrow \infty$ or $r\rightarrow 0$.

Our experiment for the distillation of corrupted squeezed states consists of three parts (Fig.~\ref{setup}): the preparation, distillation and verification. We observe the squeezing of sidebands at $17.5\pm0.5$~MHz relative to the optical carrier frequency to avoid low frequency technical noise. The preparation of the mixed state of Eq.~\ref{W} is accomplished by combining a squeezer with a controllable noise source. We use a polarization squeezer exploiting the Kerr nonlinearity experienced by ultrashort laser pulses in optical fibers~\cite{heersink05.ol}. Using a birefringent fiber, two quadrature squeezed states can be simultaneously and independently generated. Stable overlap of these pulses allows us to generate Stokes parameter squeezing. Considering the Stokes plane orthogonal to the classical excitation ($S_3$, circularly polarized), the 'dark' plane ($S_1-S_2$), it is found that the polarization squeezing observed in this mode is mathematically equivalent to quadrature vacuum squeezing~\cite{josse03.prl,heersink05.ol}. We treat the two synonymously. The classical excitation can then be thought of as a perfectly matched local oscillator. From this source we observed $\Delta^2 X_{sq} =-3.1 \pm 0.3$~dB relative to the quantum noise level. The anti-squeezed quadrature contains the large excess phase noise characteristic of pulse propagation in glass fibers, here $\Delta^2 P_{sq} = +27 \pm 0.3$~dB. These noise signals are observed using balanced detector pairs with 85\% quantum efficiency.

\begin{figure}
\includegraphics[scale=0.4]{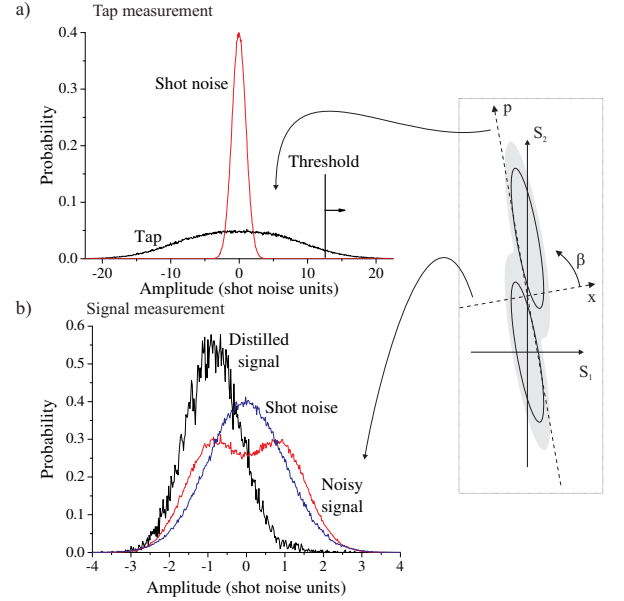}
\caption{Experimentally measured marginal distributions, centered at zero for convenience, outlining the distillation of a squeezed state from a non-Gaussian mixture of squeezed states; a) tap measurement $p_t$, b) signal measurement $x_s$. Inset: Phase space representation of the mixed state and the projection axes used in the measurements.}
\label{marginals}
\end{figure}

The non-Gaussian noise source is implemented by executing a fixed phase space displacement of the squeezed state with a probability $\gamma =0.5$. The displacement is generated by a phase modulation in one of the linear polarization modes at the fiber input using an electro-optic modulator at 17.5~MHz. This produces a corresponding displacement along the $S_2$ polarization after the fiber (Fig.~\ref{marginals}, inset). The modulation depth governs the amount of phase space displacement. By periodically switching the modulation on and off, the displacement is toggled from maximum to zero at a frequency of 500~kHz.

This non-Gaussian state, with $\Delta^2 X =+1.4 \pm 0.3$~dB, is fed into the distiller. It consists of two operations: i) the tap measurement of a certain quadrature on a small portion of the beam; ii) the signal post selection conditioned on the tap measurement. The latter could be implemented electro-optically to probabilistically generate a freely propagating distilled signal state. To avoid such complications our conditioning is instead based on data post selection using a verification measurement. The tap and the signal are recorded simultaneously, yielding data pairs, and the signal is selected dependent on the tap value. These measurements are implemented as Stokes measurements in the 'dark' plane. For a circularly polarized beam the rotation of a half-wave plate introduces a relative phase shift between the right-hand circular polarization (squeezed state) and the left-hand polarization (local oscillator) when observing the difference signal after a polarization beam splitter~\cite{heersink05.ol}. This is equivalent to the phase scan of a homodyne detector measuring arbitrary quadratures in the 'dark' plane. 

\begin{figure}
\includegraphics[scale=0.8]{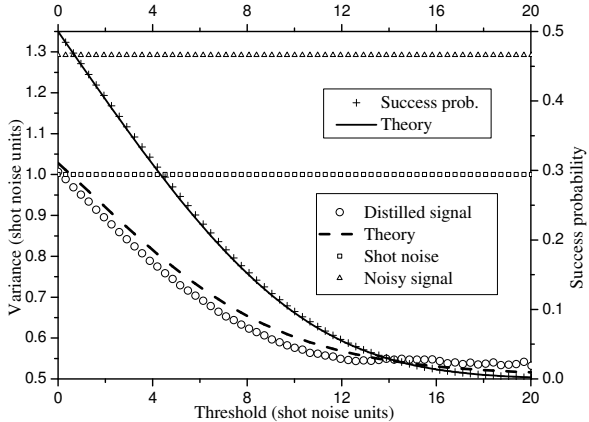}
\caption{Experimentally and theoretically distilled squeezing (left) and success probability (right) as a function of post selection threshold for two displacements. The threshold is given relative the center of the marginal distributions.}
\label{threshold}
\end{figure}

The RF photocurrents of the photodetectors are mixed with an electronic local oscillator at 17.5~MHz and digitized with a fast AD converter at $10^7$~samples per second with a 16~bit resolution. We define our state of the electro-magnetic field to be a time window of 1~$\mu$s. By digital filtering and averaging over time bins of 1~$\mu$s we derive a photocurrent value for each bin. In this process the 1~$\mu$s time bins of our signal are synchronized to the modulator switching period, such that each bin is recorded entirely during an 'on' or an 'off' period. Thus by measuring the anti-squeezed quadrature in the tap on an ensemble of identically prepared noisy states we construct the distributions in Fig.~\ref{marginals}(a). The simultaneous measurement of the signal beam recorded the orthogonal, squeezed quadrature. The modulation was chosen such that the variance of the noisy signal was just greater than that of the shot noise (Fig.~\ref{marginals}(b)). Performing post selection on this data by conditioning it on the tap measurement, we observe a recovery of the squeezing. That is, the distilled signal distribution is narrower than that of the shot noise (Fig.~\ref{marginals}(b)). We measured $\Delta^2 P_t = +17.5.0 \pm 0.3$~dB relative to the shot noise. Conditioning on the tap, the noisy signal variance, $+1.1 \pm 0.3$~dB, fell to $-2.6 \pm 0.3$~dB after distillation.

Using the data shown in Fig.~\ref{marginals}, the distilled signal variance was investigated as a function of the post selection threshold. In Fig.~\ref{threshold} we notice that an increasing threshold decreases the signal variance, ultimately approaching the input squeezing. This agrees well with the exponential increase in squeezing predicted by Eq.~\ref{distill}, given by the dashed line. As the threshold increases, the success probability or amount of distilled data decreases to zero causing an increase in the statistical error on the variance. Thus a compromise between the post selected variance and probability of success must be made.

\begin{figure}
\includegraphics[scale=0.4]{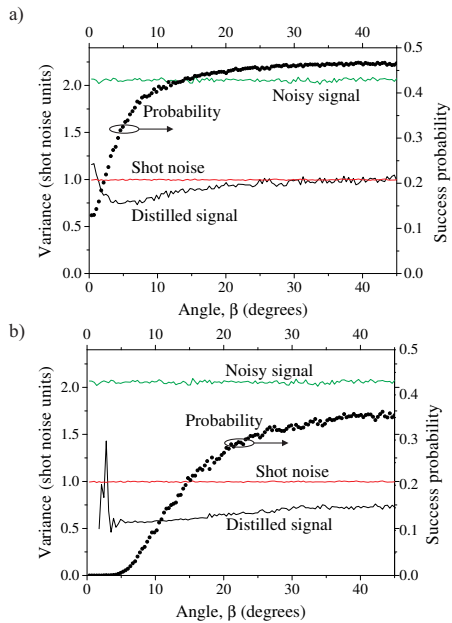}
\caption{Distilled variance (left axis) and success probability (right axis) as a function of the quadrature angle relative to the squeezed quadrature (angle $\beta$ in Fig.~\ref{marginals}) in the tap measurement for a post selection threshold of a) 1.3 and of b) 5.3 shot noise units.}
\label{tapturn}
\end{figure}

The effectiveness of a given threshold depends on i) the projection of the displacement onto the measured quadrature and ii) the variance of the measured quadrature. Fig.~\ref{tapturn}(a) and (b), each with a different threshold, shows this effect. The measured tap quadrature was rotated by an angle $\beta$ (see Fig.~\ref{marginals}), effectively changing the displacement size. We observe the best distillation for small angles where the displacement ($\bar{x}_0-\bar{x}_1$) to threshold ($p_t$) difference is largest. It is seen that the quality of the distillation decreases with increasing $\beta$ as the projection of the displacement onto the measured quadrature increases. We note however that for large thresholds the distillation quality is approximately independent of $\beta$ or the measured quadrature.

\begin{figure}
\includegraphics[scale=0.4]{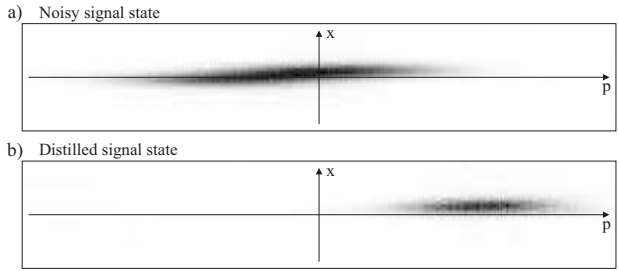}
\caption{Density plots of the Wigner function distributions for a) the non-Gaussian mixed state measured at the verification setup and b) the Wigner function of the corresponding distilled state. Note: To aid visualization the plots have been vertically rescaled by a factor of two.}
\label{wigner}
\end{figure}

We also measured the Wigner function, for the first time in fiber-based systems, of both the mixed and the distilled states. Rotating the half-wave plate in the verifier (Fig.~\ref{setup}) allows observation of all the squeezed state's quadratures for a constant tap measurement. We made 128 equally spaced projections in phase space, each of $3.5\cdot 10^6$ data points to which we applied the inverse Radon transformation~\cite{toft96.phd} to derive the Wigner functions. Fig.~\ref{wigner}(a) shows the density plots of the Wigner function associated with the noisy state; its non-Gaussian nature is evident. In Fig.~\ref{wigner}(b) we present the Wigner function of the post selected signal and observe a distribution very similar to that of a single squeezed state. Thus the purity of the state relative to the non-Gaussian state is increased and a corresponding recovery of the squeezing is seen. The shift to the right reflects the post selection process as well as the renormalization of the distilled data. Further investigation of the Wigner functions of fiber-based squeezed states is described elsewhere~\cite{marquardt06.tbp}.

We have sucessfully experimentally demonstrated the probabilistic distillation of continuous variable non-classical states from a non-Gaussian mixture of squeezed states. This was accomplished by the thorough investigation of a specific source of non-Gaussian noise, namely a linear shift in phase space. The methods presented here can be implemented for many other forms of non-Gaussian noise, e.g. phase space rotations, which will be the subject of further experiments. Another extension of this work presented would be to perform a conditional optical operation, i.e. phase shift or displacement, on the signal beam. Thus not only the distillation demonstrated here but also a purification of the excess noise of the initially squeezed states could be implemented, generating an even purer non-classical resource than produced here. Whilst we have focused on single-mode squeezed states, these techniques can assuredly be extended to two-mode squeezed systems. This means that continuous variable entanglement distillation is possible using local Gaussian operations and classical communication if the two-mode squeezing is corrupted by non-Gaussian noise.

We thank M.~Chekhova for fruitful collaboration. This work has been supported by the EU project COVAQIAL (project no. FP6-511004). R.~F. was supported by: 202/03/D239 of GACR, MSM6198959213 of MSMT CR and by the Alexander von Humboldt fundation.


\begin{thebibliography}{99}
\bibitem{note} We have choosen to consider the phase quadrature for mathematical convenience. A thorough analysis of the optimal quadrature is left to a future article.
\end{thebibliography}
\end{document}